\documentclass[10pt]{article}
\usepackage{amsmath}
\usepackage{amssymb}
\usepackage{amscd}
\usepackage{bm}
\usepackage{stmaryrd}
\usepackage{bbold}
\usepackage{eufrak}
\usepackage{isomath}
\usepackage{theorem}
\usepackage{xspace}
\def\J{{\cal J}}

\def\r{{\mathcal{R}}}
\def\i{{\mathcal{I}}}
\def\j{{\mathcal{C}}}

\def\Z{\mathbb Z}
\def\R{\mathbb R}
\def\C{\mathbb C}

\def\JJ{\mathfrak J}

\def \l0{l_0}
\def\Tau{\tau}
\def\heis{\mathfrak{h}}
\def\z{\mathfrak z}
\def\demi{{\frac{1}{2}}}

\newcommand{\beq}{\begin{equation}}
\newcommand{\eeq}{\end{equation}}
\newcommand{\ba}{\begin{array}}
\newcommand{\ea}{\end{array}}
\newcommand{\beqa}{\begin{eqnarray}}
\newcommand{\eeqa}{\end{eqnarray}}

\newtheorem{deff}[subsection]{Definition}
\newtheorem{lemma}[subsection]{Lemma}
\newtheorem{prop}[subsection]{Proposition}

\begin{document}

\begin{titlepage}
\title{Bloch waves  and  \\  Non-commutative Tori of Magnetic Translations }
\author{Tekin Dereli\footnote{E.mail:tdereli@ku.edu.tr} \\
{\small Department of Physics, Ko\c{c} University, 34450 Sariyer-\.{I}stanbul, Turkey}\\  \\
Todor Popov \footnote{E.mail:tpopov@aubg.edu}\\
 {\small Institute of Nuclear Research and Nuclear Energy,} \\ 
{\small Bulgarian Academy of Sciences,}  \\{ \small bld. Tzarigradsko chauss\'ee, 72,  Sofia 1784, Bulgaria} \\
{\& \small American University in Bulgaria} } 
\maketitle

\abstract{ \noindent We review the Landau problem
of an electron in a constant uniform magnetic field.
The magnetic translations are the invariant transformations
of the free Hamiltonian.
%\marginpar{abstract is modified} 
A K\"ahler polarization of the plane has been used for the geometric quantization. 
Under the assumption of quasi-periodicity of the wavefunction the  magnetic translations in the Bravais lattice generate 
a non-commutative quantum torus. We concentrate on
the case when the magnetic flux density is a rational number. 
The Bloch wavefunctions 
 form a finite-dimensional  module of the noncommutative torus
of magnetic translations  as well as of its commutant
which is
the non-commutative
torus of magnetic translation in the dual Bravais lattice.
The bi-module structure of the Bloch waves is shown to be the connecting link
   between  two Morita equivalent non-commutative tori.

 %The modular parameter $\tau$ depends on the magnetic field.
 %Taking the spin of the electron into account we are naturally lead to work with supersymmetric extension of the magnetic translation
%group and to
 %supersymmetric quantum theta functions. The latter objects (introduced by Yuri Manin)  span  bimodules of the magnetic supertranslations
%and provide a nice physical exemple of Morita equivalence between quantum super-tori. 
%Potential applications to the Quantum Hall Effect are discussed.
}

\end{titlepage}

\section{Introduction}

A magnetic field  modifies the geometry of space in the sense that
 the geodesic motion of a free electron in a constant uniform magnetic field is no longer a straight line
but a helix. In the center of mass system the trajectory
is reduced  to a circular orbit in the plane perpendicular to the magnetic field.
A  quantum mechanical description of the electron's geodesic motion in the plane is given by  coherent states
through geometric quantization while
 their internal symmetry is captured by Zak's  magnetic translations \cite{Zak}. K\"ahler manifolds  are especially 
suitable for geometric quantization, their holomorphic
sections  provide   the Bargman-Fock Hilbert space. 
The different complex structures give different quantizations.
However, on the plane of motion these are all equivalent due to an intertwining action of the metaplectic group $Mp(2,\R)$
that interpolates between Heisenberg representations (squeezing operator). When imposing periodic
boundary conditions on the plane, the electron motion (Bloch waves)
is described by the holomorphic sections on the torus. 
The space of
sections depends on the choice of the complex structure 
on the torus specified by a modular parameter $\tau$. On the torus 
different values of the modular parameter can give physically equivalent theories 
which is an example of duality. 
The rational values of the magnetic flux density correspond
the rational values of the modular parameter, upon quantization these
 lead  to  a splitting of the Hilbert space  into disjoint subspaces of Landau levels. 
The lowest Landau level  that is the Hilbert subspace
of  degenerated ground states of the theory is endowed  
 with a matrix action of the magnetic translation operators. 

 The Landau problem on a torus is equivalent to the topological sector 
of the Maxwell-Chern-Simons theory coupled with matter describing the low energy effective theory
of the Fractional Quantum Hall  (FQH) effect \cite{Wen, Kogan, Ho}. %\marginpar{ref \cite{Kogan} added}
 The structure of the degeneracy of the ground states in FQH effect is a major problem 
in condensed matter physics \cite{Wen} related to the modular transformations
of the Hall states.

 The phyical $T$-duality interpolates between the algebra of the magnetic translations in
Bravais lattice and its commutant algebra with translations in the dual Bravais lattice. In the mathematical
parlance, we are dealing with a pair of Morita equivalent non-commutative tori leading to  equivalent physical theories
\cite{Schwarz98}.
% Stone-von Neumann  theorem garantees
 %the uniqueness of these modules.

The remainder of the paper is organized as follows:
In Section 2 we review the Landau problem on the plane in terms of 
holomorphic $z$ and anti-holomorphic $\bar{z}$ coordinates and derivatives.
In Section 3 we endow the phase space with a structure of a K\"ahler 
manifold and consider its quantization. The generic complex structure is parametrized by a modular parameter $\tau$. In Section 4 we introduce the non-commutative algebra
of Zak's magnetic translations \cite{Zak}.
In Section 5 we introduce a periodic potential  and compactify the configuration space  to a torus:
the complex plane modulo the Bravais lattice $\Lambda$. The magnetic translations
along $\Lambda$ define a non-commutative torus $\mathbb T_\kappa$, while its commutant algebra
$(\mathbb T_\kappa)^\prime$ spanned by the magnetic translations in the dual Bravais 
lattice $\Lambda^\ast$
turns out to be isomorphic to a non-commutative torus $\mathbb T_{1/\kappa}$ for a different
value of the magnetic flux parameter $\kappa^\prime=1/ \kappa$.
In Section 6 we choose the magnetic flux to be given by a rational number $\kappa=\frac{M}{N}$
and consider the magnetic translation eigenvalue problem. Then the lowest Landau level
eigenfunctions are parametrized by the factor-lattice $$\Lambda^\ast/\Lambda \cong \mathbb Z_M \times \mathbb Z_N \ .$$ In Section  7 we treat the problem of modular invariance and propose a modular invariant partition function. Section 8 is about the
squeezing operators, that realize the metaplectic group action
 intertwining different complex structures.
In Section 9 we show that the Bloch waves in
any Landau level form a matrix 
with a left action of the magnetic translation torus $\mathbb T_\kappa$
and a right action of the dual magnetic translation torus $\mathbb T_{1/\kappa}$.
The  $(\mathbb T_\kappa,\mathbb T_{1/\kappa})$-bimodule guarantees 
the Morita equivalence of the two non-commutative tori.
Section 10  consists of a conclusion and perspectives for future work.

 %We obtain a supersymmetric extension of the Stone- von Neumann theorem.

%\noindent Magnetic units: 

\section{ Landau levels and their degeneracy}

The classical  motion of an electron of mass $m$ and charge $e$ in an uniform magnetic field $\vec B$
is the Larmour (cyclotron) motion   with constant velocity
along the magnetic field $\vec B$  and circular  motion
 in the system of center of mass. 
We choose the $\hat{z}$-axis  along the magnetic field,
so $B_i=(0,0,B)$ and assume $B>0$, {\it i.e.,} the 
vector $\vec B$ is pointing upwards.
In this way the dynamics is reduced to a rotation in a $(x,y)$-plane 
%$(x,y)$
with the cyclotron frequency $$\omega=\frac{|eB|}{mc}$$
where the direction of the rotation of a particle with negative charge $-e$ is anti-clockwise.
The magnetic field introduces also a length scale, {\it the magnetic length},given by
$$
l_B= \sqrt{\frac{\hbar c}{e B}} \ .
$$
The  flux $BA$ of the magnetic field $B$ through an area $A$ is quantified
in 
units of elementary magnetic flux $\frac{hc}{e}$. We let
\[
\frac{BA}{{hc}/{e}}=\frac{BA}{2 \pi{\hbar c}/{e}}= \frac {A}{2 \pi l_B^2} = \kappa \ . \]
The number $\kappa$ is  the density of the magnetic field lines
(quanta of the flux).
%(or the density of states).
%The kinetic momenta $\bm{P}$ incorporates the

The quantum mechanical description of the
cyclotron motion of an electron is first studied by Landau
\footnote{For an concise introduction to Landau problem we send the reader to the
inspiring lectures 
\cite{Murayama}}.
  The Hamiltonian of the Landau problem 
is defined by the minimal coupling of the electron to an external magnetic field described by the 
electromagnetic vector potential $\bm{A}$:
\beq
\label{Ham}
H= \frac{1}{2m}\left( \bm{p} - \frac{e}{c}\bm{A} \right)^2
=:\frac{1}{2m} \bm{P}^2 \ .
\eeq
The constant uniform magnetic field $\bm B= B \hat{z}$ along the $z$-direction
can be obtained from different potentials $\bm{A}= (A_x, A_y)$
in the plane.  One particular choice is the symmetric gauge
\beq
\label{symg}
\bm A = (A_x, A_y)= \frac{B}{2} (-y,x) \ ,  \qquad A_i=-\frac{B}{2}
\epsilon_{ij}x^j \ .
\eeq
The coordinates on the plane $(x,y)$ and the canonical momenta $ \bm p =(p_x,p_y)
= -i \hbar (\partial/\partial x, \partial/\partial y)$  provide the standard (so called Schr\"odinger) representation
of the Heisenberg commutation relations
\[ [p_k, p_l]=0, \qquad [x_k, x_l]=0, \qquad [p_k,x_l]=-i\hbar \delta_{kl} \  . \]
However,  other natural coordinates on the phase space
$(x,y,p_x,p_y)$ may reflect better the change of the geometry due to the
magnetic field, namely,  the kinetic  momenta  \footnote{We suppose that we have a positive metric $g_{ij}=\delta_{ij}$, hence
 $P_i=P^i$ }  $ \bm P =(P_1,P_2):=(P_x,P_y)$ and the coordinates of the center of mass $\bm X= (X_1,X_2):= (X, Y)$  that are given by
\[
P_i = m\dot{x}_i
= m \frac{\partial H}{\partial p_i}=
 %\frac{\partial L}{\partial x_i}=
 p_i - \frac{e}{c}A_i, 
\qquad
X_i = x_i + \frac{1}{m \omega} \epsilon_{ij} P^j \ .
\]
The center of mass coordinates $\bm X= (X, Y)$
 are the coordinates of the center of the cyclotron motion.
These  are  integrals of motion 
$ {\dot{X}}=0= \dot{Y}$ and as such decouple from the  system.
Therefore the  observables on the phase space split into two
independent commuting algebras
\[
\qquad [P^i,X_j] = 0 \ ,
\]
\[
[P_x, P_y]= i\frac{\hbar e}{c}B =i \hbar m \omega=\frac{i\hbar^2}{l_B^2}\ ,
\qquad \qquad \qquad \qquad [X,Y]= -i \frac{\hbar}{m \omega} = -i l^2_B \ . \]
Since the ``center of mass'' coordinates $(X,Y)$ commute{\footnote{This commutation is specific for the symmetric gauge.}} with  the Hamiltonian (\ref{Ham}) they are also referred to as ``zero modes''.
The presence of the magnetic field $\pm B$ leads to a  deformation of 
 the commutators 
between the kinetic momenta $P_x$ and $P_y$ and similarly between the center of mass coordinates $X$ and $Y$.
Geometrically,  the magnetic field $\pm B$ is the curvature of
a connection $\bm P$.

{\bf Landau levels.} In view of their noncommutativity,
the operators $P_x$ and  $P_y$ alone can be thought
as canonically conjugated operators in a reduced phase space.
Their subalgebra is quantized in terms of creation and annihilation
operators\footnote{Our conventions slightly differ from the standard $a^{\pm}=\frac{1}{\sqrt{ 2 \hbar m \omega}} (P_x \mp i P_y)$.}
\[
a^{\pm}=\frac{1}{\sqrt{ 2 \hbar m \omega}} ( - P_y \mp i P_x) \ .
%\qquad [a^-, a^+]=1
%\frac{1}{{ 2 \hbar m \omega}}[P_x +i P_y,P_x- iP_y]=1
\]
Indeed it is straightforward to check that $[a^-, a^+]=1$.
%\[
%[a^-, a^+]=\frac{1}{{ 2 \hbar m \omega}}[i P_x - P_y,-i P_x- P_y]=1
%\]
The operators $a^+$ and $a^-$ are the raising and lowering operators
for the eigenstates of the Hamiltonian (\ref{Ham}) which may also be given as
%, which are the so called Landau levels
\[
H= \hbar \omega \left(a^+ a^- + \frac{1}{2}\right)
= \hbar \omega\frac{1}{2} \{a^+, a^-  \} , 
\qquad  [H, a^\pm]=\pm a^\pm \ .
\]
Hence we conclude that  the Landau problem is equivalent to 
a simple harmonic 
oscillator  problem %where $\omega$ is 
 with the cyclotron frequency $\omega$.

 The Landau levels  are the eigenspaces of the Hamiltonian
$H$ with energies 
\[
E= \hbar \omega \left(n + \frac{1}{2}\right) ,
\qquad n=0,1,2, \ldots  , 
\]
depending on the quantum  number operator $n$,
that  counts the quanta of the magnetic field $B$
transversal to the plane of motion. % magnetic field  .
The lowest Landau level is the ground state of the system
and it satisfies the equation
\[
a^{-} \Phi_{(0)} = 0 \ .
\]
All other Landau levels are excited states over the ground state $\Phi_{(0)}$.

{\bf Complex coordinates and holomorphic sections.} 
Let us introduce complex holomorphic and antiholomorphic coordinates on our  $(x,y)$-plane:
\[z= (x+ iy)/l_B  , \qquad \bar{z} = (x- iy)/l_B . \]
The respective derivatives
obeying $\partial z=1=\bar{\partial} \bar{z}$
and $\partial \bar{z}=0=\bar{\partial}z$ are explicitly given by
\[
 \partial =\frac{\partial}{\partial z}= \frac{l_B}{2}\left(\frac{\partial}{\partial x}
-i \frac{\partial}{\partial y}\right) \ , \quad \bar{\partial}=
\frac{\partial }{\partial\bar{z}}= \frac{l_B}{2}\left(\frac{\partial}{\partial x}
+i \frac{\partial}{\partial y}\right) \ .
\]
In terms of these  complex coordinates, the creation and   annihilation 
operators 
$a^\pm $ %=\left( \mp i P_x  -P_y \right)/ \sqrt{2 \hbar m\omega} $
 are represented by the succinct expressions 
\[
a^-= %-i 
\sqrt{2}\left( \bar{\partial}+ \frac{z}{4 }
\right) \ , \qquad 
a^+=%-i 
- \sqrt{2}\left(
{  {\partial}- \frac{\bar{z}}{4  }}
\right) \ .
\]
Here we  used the vector potential $\bm A$ in the symmetric gauge (\ref{symg}).

The following differential equation for the ground state in complex coordinates 
\[
a^-\Phi_{(0)}(z, \bar{z})=
%-i 
\sqrt{2}\left(
  \bar{\partial}+ \frac{z}{4 }
\right)\Phi_{(0)}(z, \bar{z})=0
\]
is now
easy to solve, providing the {\it holomorphic sections}
of a line bundle with connection $\bm P$. Its solution is the Gaussian factor multiplying
an  {\it arbitrary} holomorphic function $f(z)$,
\beq
\label{ground}
\Phi_{(0)}(z, \bar{z})= f(z) \exp \left(-\frac{z \bar{z}}{4} \right) \ .
\eeq
The ground state becomes infinitely degenerate due to the residual symmetry of the ``zero modes''. A non-normalized basis in the space of different
holomorphic sections (ground states) reads
\beq
\Phi_{(0),n}(z, \bar{z})= N_n z^n  \exp \left(-\frac{z \bar{z}}{4 } \right) \ , \qquad N_n^2 =1/( \pi n! 2^{n+1})
\eeq
%corresponding to 
given by  the monomial basis $\{z^n\}_{n \geq 0}$ of (polynomial) holomorphic functions $f(z)$.
The space of holomorphic states has a positive definite norm
\[
\langle \Phi_{(0)}%(z, \bar{z})
|\Psi_{(0)}%(z, \bar{z})
 \rangle
= \int (-i)d z \wedge d\bar{z} \, \exp \left(-\frac{z \bar{z}}{2 } \right) f(\bar{z})g(z)
\]
where the non-holomorphic part 
yields a measure on the 
space of holomorphic functions $f(z)$.

{\bf Center of mass coordinates $X$ and $Y$.}
The degeneracy of a Landau level can be removed if
we consider the zero modes $\bm X$ as internal degrees of freedom.
We quantize the center of mass coordinates $X$ and $Y$ by
the creation $b^+$ and annihilation $b^-$  operators that are given by
\[
b^{\pm} = \frac{1}{  l_B\sqrt{2}} (X\pm i Y) \ ,
\qquad \qquad [b^-,b^+]=1 \ .
\]
The operators $b^{\pm}$ commute with the $a^{\pm}$
since these  are made out of zero modes.

In complex coordinates the ``zero mode" operators  are given by
the concise expressions %looks like
\[
b^-=  \sqrt{2}\left(
{  {\partial}+ \frac{\bar{z}}{4 }}
\right) \ ,\qquad 
b^+
=-\sqrt{2}\left(
{  \bar{\partial} - \frac{z}{4 }} 
\right) \, \ .
\]
It is worth noting that expressions for $b^\pm$ are 
mapped on to the ones for $a^\pm$ by complex conjugation $z \rightarrow \bar{z}$.

The operators $b^{+}$ and $b^-$ raise and lower the
angular momentum such that
 the state
$\Phi_{(0),0}(z, \bar{z})$  has minimal angular momentum :%index of the  ground states $\Phi_n(z, \bar{z})$.
\[
b^- \Phi_{(0),0}(z, \bar{z})=b^- N_0   exp \left(-\frac{z \bar{z}}{4 } \right)= N_0  \sqrt{2}\left(
{  {\partial}+ \frac{\bar{z}}{4 }}
\right)   exp \left(-\frac{z \bar{z}}{4 } \right) =0 \ .
\]
From this minimal state we can obtain all other states
in the lowest Landau level,
\[
(b^+)^n \Phi_{(0),0}(z, \bar{z})=
(b^+)^n  exp \left(-\frac{z \bar{z}}{4} \right)
= \sqrt{n!} N_n z^n   
exp \left(-\frac{z \bar{z}}{4} \right) = \sqrt{n!}\Phi_{(0),n}(z, \bar{z}) \ .
\]

\section{ Quantization and K\"ahler polarization.} 
A K\"ahler manifold is a  complex manifold equipped with a non-degenerate Hermitean form whose real part is a metric (symmetric) form and whose imaginary part is a symplectic (antisymmetric) form. Complex manifolds can be seen
as an even-dimensional real manifold provided with $\J^2=-1\!\!1$. There are  three mutually compatible structures on a K\"ahler manifold, namely, the metric, symplectic and 
complex structures
it is sufficient to specify any two of these structures in order to obtain the third one. 
%and thus the hermitean Ka\"hler form.

The general  holomorphic and antiholomorphic coordinates on the complex plane are introduced by the following mappings from  
$\R^2$ to $\C$:
\beq
\label{z}
z = (x + \tau y)/l_B , \qquad \bar{z} = (x + \bar{\tau} y)/l_B , \qquad  Im \, \tau >0 \ 
\eeq
where $\tau=\tau_x+ i\tau_y$ is a complex number. The  magnetic length $l_B$ in the  denominator  fixes the unit of length.  It  makes  the complex variable $z$ dimensionless
and simplifies some formulas. The condition
$Im \, \tau >0 $ implies that the parallelogram is not degenerate
and its area is oriented.

The choice of holomorphic $z$ and antiholomorphic $\bar{z}$ coordinates given by Eq. (\ref{z}) is equivalent to  the choice of a  complex structure $\J=\J(\tau)$ on the $(x,y)$-plane:
the eigenspaces of the operator $\J$ being the holo\-mor\-phic and antiholo\-mor\-phic coordinates
\[
\J \, z = -i z  \ , \qquad \qquad \J \, \bar{z} = i \bar{z} \ .
\]
The standard holomorphic  
$z=x+iy$ and antiholomorphic $\bar{z}=x-iy$ 
coordinates 
( $\tau=i$) correspond to the  standard complex structure 
\[
\J_{0}= \left(\ba{rr} 0 & 1\\-1 & 0  \ea\right) \ .
\]
%It is achieved at the special  point $\tau=i$ in Eqs. (\ref{z}).
The general complex structure $\J$ is found from the linear system
\beq
\J(x + \tau y) = - i (x + \tau y) \ ,
\qquad 
\J (x+ \bar{\tau} y) = i (x+ \bar{\tau} y) \ .
\eeq
Solving it for $\J(x)$ and $\J(y)$ we get the matrix 
%in $SL(2,\R)$ 
of the operator $\J$:
\beq
\label{jj}
\J % \left(\ba{c} x \\ y \ea \right)
=\frac{1}{ Im \, \tau}
\left(\ba{cc} Re \, \tau & |\tau|^2\\ -1 & -Re\, \tau \ea \right)
%\left(\ba{c} x \\ y \ea \right)
\in SL(2,\R) \ .
\eeq

The metric on a K\"ahler space satisfies
the compatibility condition $g(x,y)=\Omega(\J x,y)$.
Hence from the canonical symplectic form $\Omega=dx \wedge dy$ 
and the complex structure  $\J$ given by Eq.(\ref{jj}),
one gets the metric  
\[
g = \J^t \left(\ba{rr} 0 & 1\\-1 & 0  \ea\right) =
\frac{1}{ Im \, \tau}
\left(\ba{cc} 1 & Re \, \tau \\  Re\, \tau &|\tau|^2  \ea \right) \ .
\]
The condition $Im\, \tau >0$ guarantees the positivity of the metric.

In our Landau problem we have tacitly introduced a K\"ahler manifold starting with a symplectic
manifold $M\cong \R^4$ with coordinates $(p_x,p_y,x,y)$  with the 2-form $\Omega= dp_x \wedge dx + dp_y\wedge dy$, and then choosing a standard complex structure 
%$\J=\J_0$ 
given by the $4\times 4$ matrix
\[
 \left(\ba{rr} {\bf 0} & 1\!\!1 \\-1\!\!1 & {\bf 0}  \ea\right) \ .
\]
This is 
 a K\"ahler polarization which turns $(M, \Omega)$ into a
K\"ahler manifold. When we consider the lowest Landau level
 we reduce our phase space to 
a two dimensional one.

Having introduced  general holomorphic and antiholomorphic coordinates
 (thus a complex structure
$\J(\tau)$) we proceed by defining the holomorphic and antiholomorphic derivatives
\[
\partial =
\frac{\partial}{\partial z} = \frac{l_B}{\tau - \bar{\tau}}
\left( - \bar{\tau}\frac{\partial}{\partial x}+ \frac{\partial}{\partial y} \right) \ ,
\qquad 
\bar{\partial}=
\frac{\partial}{\partial \bar{z}} = \frac{l_B}{\tau - \bar{\tau}}
\left( {\tau}\frac{\partial}{\partial x}- \frac{\partial}{\partial y} \right) \ .
\]
such that $\partial z=1=\bar{\partial} \bar{z}$
and $\partial \bar{z}=0=\bar{\partial}z$ hold true (compatible with Eq. (\ref{z})).

Next step is the quantization of the K\"ahler manifold
$(M, \Omega, \J)$:\footnote{For expository notes on quantization on K\"ahler manifolds see \cite{Todorov}.} one  defines the creation and annihilation operators depending  on the complex structure $\J(\tau)$.
Mimicking the standard case $\tau=i$, we define
the algebra of  raising and lowering operators $a^{\pm}_\tau$
changing the Landau level
\[
a^{+}_\tau=\frac{(\tau P_x -  P_y)l_B}{\sqrt{2\, Im \, \tau }\sqrt{\hbar}} \ ,\qquad \qquad
a^{-}_\tau= \frac{(\bar{\tau} P_x - P_y)l_B}{\sqrt{2\, Im \, \tau }\sqrt{\hbar}} \ , \qquad \qquad [a^{-}_\tau, a^{+}_\tau] =1\]
and its commutant algebra of ``zero mode'' operators $b^{\pm}_\tau$
\[
b^{+}_\tau= \frac{X +{ \tau} Y}{l_B\sqrt{2\, Im \, \tau }} \ ,
\qquad \qquad
b^{-}_\tau= \frac{X +\bar{ \tau} Y}{l_B\sqrt{2\, Im \, \tau }} \ ,
\qquad \qquad [b^{-}_\tau, b^{+}_\tau] =1 \ .
\]
Translated into complex coordinates these operators read
\beqa
\label{aa}
a^+_{\tau}= -\sqrt{2 \, Im \,  \tau}
 \left(\frac{\partial}{\partial {z}}- \bar{z}/{(4\, Im\, \tau)}  \right) \ ,\qquad 
a^-_{\tau}=  \sqrt{2\, Im \, \tau} \left(\frac{\partial}{\partial\bar{z}} + {z} /(4\, Im \, \tau)  \right) ,
\\
\label{bb}
b^+_{\tau}= -\sqrt{2 \, Im \,  \tau} \left(\frac{\partial}{\partial \bar{z}} - z/{(4\, Im\, \tau)}  \right) \ , \qquad 
b^-_{\tau}=  \sqrt{2\,  Im \, \tau} \left(\frac{\partial}{\partial z} + \bar{z} /(4\, Im \, \tau)  \right) \ .
\eeqa

The ground states $\Phi_{0,\tau}$ (\ref{ground}) in the lowest Landau level
are annihilated by the lowering operator $a^-_\tau$ ($\Phi_{0,\tau}$ are covariantly constant sections):
\[
a^-_\tau \Phi_{0,\tau} =0 \ .
\]
The solutions turn out to be  Gaussians centered around the origin $z=0$,
\[
\Phi_{0,\tau}  = C \exp \left(-\frac{ z \bar{z} }{4 \, Im \,\tau }\right)
\]
where the spread of the ``bell'' depends on $Im\, \tau$.
The function $\Phi_{0,\tau}$ yields a {\it coherent state},
that is, 
a state that saturates the lower bound in the Heisenberg uncertainty inequalities.

More generally a coherent state centered around the 
 point $z=z_0 $ satisfies
\[
a^-_\tau \Phi_{z_0,\tau} = z_0 \Phi_{z_0,\tau} \qquad
\Phi_{z_0,\tau}  = C \exp \left(-\frac{|z-z_0|^2 }{4 \, Im \, \tau }\right) \ 
\]

In fact, all the ground states $\Phi_{z_0}$ are {\it coherent states}. % minimizing the Heisenberg uncertainty inequalities. 
Indeed the position operator expectation value yields
\[
\langle \Phi_{z_0, \tau}| z | \Phi_{z_0, \tau} \rangle = z_0 \ .
\]

\section{Magnetic Translations }
Each Landau level is infinitely degenerate in energy. 
 The center of mass coordinates $X$ and $Y$ give rise to
 infinitesimal translation operators $\bm{T}=(T_x, T_y)$ that after rescaling 
%(with dimension of inverse length)
\[T_i = m \omega \epsilon_{ij} X^j \ ,
\qquad [T_x, T_y ] = - i m \omega \hbar \ .
\] 
The translation operators $(T_x, T_y)$ 
 commute with the momentum  $\bm{P}$  and consequently with 
the creation and annihilation operators $a^{\pm}$ and
the Hamiltonian $H$. These do not alter any Landau level, hence their name ``zero modes''.
The generators $T_i$ are similar to the momentum operators $P_i$
\[
T_i = -p_i - \frac{e}{c} A_i
\qquad \qquad
P_i= p_i - \frac{e}{c} A_i\ \qquad \qquad [T_i,P_j]=0 \ .
\]
The operators $T_i$ are the generators of  the center of mass
translations.

\begin{deff}  
Magnetic translation operator $D({\bm u })$  is a {\it finite} translation operator along a given displacement vector ${\bm u }$
\beq
\label{MTO}
D({\bm u })
= \exp \left(\frac{i}{\hbar } \bm{u \cdot T}\right)
\eeq
The unitary operator $D({\bm u })$ is also referred to as {\it displacement } operator.
\end{deff}

The  degeneracy of Landau levels  reflects their invariance 
 with respect to $D({\bm u })$.

The magnetic translation along a vector $\bm{u}=(u_x,u_y)$ acts on  states
of the Hilbert space $f(x,y)$ as translation of coordinates
and change of the phase (by the flux of the magnetic field
through a unit area)
\[
D({\bm u }) f(x,y)= e^{\frac{i}{\hbar }(u_x T_x + u_y T_y )}   f(x, y)=
e^{-i  \frac{ 2 \pi e}{h c}(A_x u_x+A_y u_y)  } 
f(x - u_x, y- u_y) \ .
\]
Alternatively, we write the magnetic translations
in  terms of complex null %``light cone'' 
variables $u$ and $\bar{u}$:
\beq
D({\bm u })= \exp \frac{ (ub^+ -\bar u b^-)}{{\sqrt{2}}} , 
\qquad \qquad u=({u_x +i u_y})/l_B \ .
\eeq
Then the translations along the rays $u=u^+$ and $\bar{u}=u^-$ yield
\beq
\label{Du}
D({\bm u }) f(z, \bar{z}) =
 \exp \left( -u (\partial + \frac{\bar{z}}{4}) +
\bar{u }(\bar{\partial}  + \frac{{z}}{4})  \right)
f(z, \bar{z})=% e^{({\bar{z}-z})/{4}}
e^\frac{\bar{u}z-{u}\bar{z}}{{4  }} f(z -u, \bar{z}- \bar{u})
\eeq

More generally,a magnetic translation operator in terms of % ``light-cone"
complex null holomorphic $u_{\tau}=(u_x + \tau u_y)/l_B$ and anti-holomorphic
  $\bar{u}_\tau=(u_x + \bar{\tau} u_y)/l_B$ coordinates\footnote{It is worth noting that our displacement
vector $\bm{u}$ have dimension of length while the (anti)holomorphic
coordinates $u$ and $\bar{u}$ are chosen to be dimensionless.}reads
\[
D(\bm{u}) = \exp \left( \frac{i}{l_B^2} \Omega(\bm{u}, \bm{X}) \right)=
\exp \frac{ \bar{u}_\tau b^+_{\tau}-u_\tau b^-_{\tau}}{\sqrt{2\,Im \, \tau}} ,
\qquad \J u= i u , \quad \J\bar{u}=-
i\bar{u} \ . %u_{}=(u_x + \tau u_y)/l_B \ .
\]
The magnetic translation action on states $f(z, \bar{z})$, in parallel with   Eq. (\ref{Du}),
is given by \footnote{By abuse of notation we  write $u (\bar{u})$
for the (anti)holomorphic coordinates $u_\tau (\bar{u}_\tau)$
whenever the complex structure is clear from the context.}
\beq
\label{trance}
D({\bm u }) f(z, \bar{z}) = 
\exp \frac{{\bar{u}z}-u\bar{z}}{{4 \, Im \, \tau }}\,\,
f(z -u, \bar{z}-\bar{u}) \qquad \J u= i u \quad \J\bar{u}=-
i\bar{u} \ .
\eeq
Applying the displacement operator $D({z_0 })$
to the ``centered'' ground state $\Phi_{0,\tau}$
will translate it and multiply by a factor depending on the angular
momentum:
\[ 
D({ z_0 })\Phi_{0,\tau}=
\exp  \left( \frac{{z_0\bar{z}-\bar{z}_0 z}}{{4 \, Im \, \tau }}
\right) \,  \Phi_{z_0,\tau}  \ .%\exp \left(-\frac{ z \bar{z} }{4 \, Im \,\tau }\right)
\]
Thus the ground states $\Phi_{z_0,\tau}$ are obtained as solutions
of the differential equation $(a^-_\tau -z_0 )\Phi=0$ in the symmetric gauge (\ref{symg}).
%On the other hand the solutions of $a^-\Phi=0$ in  the translational  invariant gauge gives 
% steadily moving solitons.

The symplectic form $\Omega$ is derivable from the so called
 K\"ahler potential $h(z, \bar{z})$ 
\[
h(z, \bar{z})= -\frac{z\bar{z}}{2 Im \, \tau} \ , \qquad \Omega =i \partial \bar{\partial} h \ .
\]
The Hilbert space of holomorphic sections together with
the metric will be given by
\beq
\Phi(z, \bar{z})= f(z) exp \left(-\frac{h}{2}  \right) \ ,
\qquad \qquad
||\Phi(z, \bar{z}) ||^2= \int |f(z)|^2 
exp \left(-h  \right) \Omega \ .
\eeq

\section{Quantum Tori}

Introduction of a periodic potential  adds to  our system
a new scale, namely % in the problem: 
the crystal lattice scale $l$ besides the magnetic length scale $l_B$. The structure of the Landau levels
is intimately related to the interplay between $l$ and $l_B$.

\noindent
{\bf Periodicity conditions.}
The crystal periods span a  Bravais lattice 
$\Lambda=\Z\bm{e}_1 +\Z\bm{e}_2$ with lattice vectors $(\bm{e}_1, \bm{e}_2)$.
 We add a periodic potential  $V(x,y)$ 
%=V(x+a,y)=V(x, y+b)$
to the free  Hamiltonian:  
%$H=\bm{P}^2/2m $
\beq
H^\prime=H + V(x,y) , \qquad
V(\bm{x})=V( \bm{x}+\Lambda)  \ .
\eeq
%For the generic periodic potential
 The crystal lattice periods are the vectors $\bm{e}_1:=(l,0)$ and $\bm{e}_2:=(l\tau_x, l\tau_y)$ in $(x,y)\in\R^2$ such that %plane
\[
\label{lattice}
%\qquad \mbox{i.e.,} \qquad
V(x,y)=V(x+l,y)=V(x + l\tau_x, y+l\tau_y)   \ .
\]

%\footnote{It is convenient to use the complex parameter $\tau=\tau_x+i\tau_x$. One has rectangular lattice when $Re \, \tau=0$
%which is cubic lattice for $Im \, \tau=1$.}
We imbed the Bravais lattice $\Lambda$ in $\C$ by $\bm{e}_1=l$ and $
\bm{e}_2=l\tau$. 
The new  periodic Hamiltonian $H^\prime$
%=H + V(x,y)$ 
becomes naturally defined %on a torus,
%after the compactification $\bm{x}\equiv \bm{x}+\Lambda$.
%In complex coordinates the configuration space of the periodic system with hamiltonian  is
on the quotient of the complex plane by the Bravais   lattice  
\[
%\z\sim \z +1 \qquad \z \sim \z +\tau \qquad
 T_{\tau} \cong \C/\Lambda= \C/(l\Z \oplus l\tau \Z) , 
\qquad \qquad \tau=\tau_x+i\tau_y
\]
which is a torus with % periods $l$ and $l\tau$ and 
complex structure $\J$ 
of modulus\footnote{The dependence on the  scale $l$ cancels in the ratio of the periods $\tau=\frac{l\tau}{l}$ of $T_\tau$.} $\tau$, $z=x+ \tau y$.

% $x\equiv x+ l \tau_x$
%and $y\equiv y+ l\tau_y$.
The translation  invariance  is broken for generic $\bm u$, 
\[
[H + V(x,y), D({\bm{u}})]\neq 0 \ .
\] However, the periodic Hamiltonian $H^\prime$ 
 commutes with a subset of the magnetic translation operators
given by
\beq
\label{MTO}
D({\bm m })= \exp \left(\frac{i}{\hbar } \bm{m \cdot \bm{T}} \right)
\ ,
\qquad \bm{m} \in \Lambda :=\{ m^1 \bm{e}_1 + m^2 \bm{e}_2
\, | \,
m^i \in \mathbb Z\} ,
\eeq
namely, the operators generating the Bravais lattice $\Lambda$. The
magnetic translations $D({\bm m })$, $\bm{m} \in \Lambda $ are the discrete symmetries of the potential $V(x,y)$.

{\bf Quantum torus.}
The elementary translations $D({{\bm e}_1})$ and $D({{\bm e}_2})$
along the lattice vectors ${\bm e}_i$ of the Bravais lattice
commute with the Hamiltonian $H^\prime$, but in general, they do not commute with each other. The lack of commutativity is due
to the flux of the magnetic field
$\bm B$  through one  plaquette of the Bravais lattice.
The magnetic flux $BA=\bm B \cdot \left(\bm e_1 \times 
\bm e_2)\right)$  through a  cell with 
area $A=|{\bm e}_1 \times{\bm e}_2|=l^2\tau_y$ in  units  of elementary magnetic flux $\frac{h c}{e}$
is the dimensionless quantity $\kappa$  given by
\beq
\label{kap}
\kappa =\frac{BA}{hc/e}= \frac{A}{2 \pi l_B^2}= \frac{Im\, \tau}{2 \pi}\frac{ l^2 }{ l_B^2} \ .
%= \frac{l_1 l_2}{2 \pi}\ , \qquad \mbox{where} 
%\qquad \qquad l_i:= \frac{L_i}{l_B}\ .
\eeq
The holonomy of the magnetic translation operators
$D({\bm u})$, Eq.  (\ref{MTO}),
 around the Bravais plaquette 
\[
D({\bm e_1}) D({\bm e_2})D({\bm e_1})^{-1} D({\bm e_2})^{-1} =e^{\frac{ie}{\hbar c}\oint \bm{A}\cdot d\bm{r}}=q
\]
is measured by the dimensionless parameter depending on 
the magnetic flux:
\[
q  = \exp \left(2 \pi i\frac{e}{hc}{\bm B \cdot (\bm e_1 \times 
\bm e_2)}\right)= \exp 2 \pi i \kappa \ .
\]
In other words the transversal (background) magnetic field $B$ induces a curvature
form $\Omega$ on the lattice $\Lambda$:
\[
\Omega({\bm m, \bm n}):= \kappa \epsilon_{ij}m^i n^j
= \kappa (\bm m \times \bm n)\cdot \bm e_3 \qquad \mbox{where} \qquad (\bm e_1 \times \bm e_2)=\bm e_3 \ .
\]
So far we have obtained a symplectic  redefinition to the magnetic translation (\ref{MTO}) in the spirit of Stokes' theorem. It
relates the circulation of the vector potential  around a loop, ({\it i.e.}, the holonomy of $\bm{A}$) to the flux of the magnetic field $\bm{B}$ through a surface spanned on that loop.
 In coordinate-free notation one has:
\beq
\label{mto2}
D({\bm m }) = \exp \left(\frac{i}{\hbar } \bm{m \cdot T}\right)=
\exp {\frac{i}{l_B^2 } dx\wedge dy (\bm{m}, \bm{X} ) } 
= \exp {{i 2 \pi}{ } \Omega( \bm{m}, \bm{X})  }
\eeq
where $\Omega$
is the   symplectic form $\Omega=\kappa \, dx\wedge dy$
of the magnetic flux  which is the product of the density
of the magnetic lines $\kappa$ and the area of an elementary cell.
Here we have bind together the symplectic structure $\Omega$
with the metric ({\it i.e.}, scalar product $\cdot$) %on the place which is typical
as usual for the K\"ahler spaces.

In the compact picture, the magnetic translations along the
periods $\bm{e}_1$ and $\bm{e}_2$ become the Wilson line operators
around the nontrivial loops of the torus $T_\tau$. Any open
path on the lattice  gives a non-trivial loop in $T_\tau$.
By construction, the magnetic flux through  a non-contractible loop is a 
non-observable quantity, being a phase $i 2 \pi \Z$. 
Thus $D(\bm{e}_i)$ are global gauge transformations.

The magnetic translation operators $D({\bm u})$
generate a {\it quantum torus} $\mathbb T_\kappa $, such that %
\[
D({\bm m}) D({\bm n}) =  
e^{i \pi \Omega({\bm m, \bm  n})}D({\bm m + \bm n})
\]
or simply in the form of Heisenberg group relation
 %\qquad
%\Rightarrow 
%\qquad
\[
 D({\bm m}) D({\bm n})=e^{i 2 \pi \kappa (\bm m \times \bm n)% \vartheta_{\bm m, \bm n}
}
D({\bm n}) D({\bm m}) \ .
\]
The phase factor measures the flux $\kappa$  of the magnetic field 
 through the parallelogram spanned by  ${\bm m}$ and ${\bm n}$.
When the flux $\kappa$ is an integer we get a commutative torus $ \mathbb C/\Lambda$.

{\bf Dual Quantum Torus.} We consider also 
  the translations in the dual lattice $\Lambda^\ast$:
\[
\widetilde{D}({\bm n^\ast}) = \exp \left(\frac{i}{\hbar }
 \bm{ n^\ast \cdot \bm{T}} \right) \ ,
\qquad {\bm{n^\ast}} \in \Lambda^{\ast} :=
\{ {n}_1{\bm e^\ast}^1 + {n}_2  {\bm e^\ast}^2 |{n}_i \in \mathbb Z\}
\] with basis
${\bm e^\ast}^i (\bm e_j)=\delta^i_j$.  The basis
${\bm e^{\ast i}}$ span the dual lattice $\Lambda^\ast$ of the Bravais lattice $\Lambda$ and  the translations $\widetilde{D}({\bm n^\ast}) $ 
are referred to as {\it dual magnetic translations}.

By construction
every magnetic translation $D({\bm m})$ commutes with every dual magnetic translation $\widetilde{D}({ \bm n^\ast})$,
\[
[{D}({\bm m}), \widetilde{D}({ \bm n^\ast})]= 0 \ .
\]
The converse is also true,  every magnetic translation, Eq. (\ref{MTO}) commuting with $D({\bm m})$ is a dual magnetic translation   $\widetilde{D}({ \bm n^\ast})$ for some $n^\ast \in \Lambda^\ast$. 

Dual magnetic translations $\widetilde{D}({ \bm n^\ast})$ generate a dual
quantum  torus $\mathbb T_\kappa^\prime \cong\mathbb T_{1/\kappa} $ 
such that
\[
\widetilde{D}({\bm m^\ast})\widetilde{D}({\bm n^\ast}) = 
e^{i \pi \widetilde{\Omega}({\bm m^\ast}, {\bm n^\ast})}
\,  \widetilde{D}({\bm m^\ast + \bm n^\ast) \ ,\qquad  
\widetilde{\Omega} ({\bm m}^\ast, {\bm n}^\ast)}
=-\frac{1}{ \kappa} \epsilon^{ij}{m}_i {n}_j \qquad 
%\tilde{\vartheta}=  -\vartheta^{-1}
 \ .
\]
The curvature  $\widetilde{\Omega}= - \Omega^{-1}$ measures the holonomy
$\tilde{q}= \exp (\frac{2\pi i }{\kappa})$ around the plaquette of the dual lattice $\Lambda^\ast$.
The dual algebra is therefore
isomorphic to a quantum torus with reciprocal value of the parameter, 
$(\mathbb T_\kappa)^\prime\cong \mathbb T_{1/\kappa} $.

It is now clear that
 the algebra $\mathbb T_\kappa$ of finite  translations $D({\bm{m}})$ in momentum space and the algebra
$(\mathbb T_\kappa)^\prime=\mathbb T_{1/\kappa}$
of dual magnetic translations $\widetilde{D}({\bm m^\ast})$ (global 
gauge transformation, winding operators) 
are commutants of each other in  a bigger algebra $\mathbb T_{\kappa} \otimes \mathbb T_{1/\kappa}$:
\beqa
\label{qtor}
D({\bm e_1}) D({\bm e_2})&= &e^{ 2 \pi i \kappa} \, \, D({\bm e_2}) 
D({\bm e_1}) \ ,\\
\widetilde{D}({\bm e_1^\ast}) \widetilde{D}({\bm e_2^\ast})&= &e^{\frac{2 \pi i}{ \kappa}} \, \,\widetilde{D}({\bm e_2^\ast})
 \widetilde{D}({\bm e_1^\ast}) \ ,\\
D({{\bm e}_i}) \widetilde{D}({\bm e_j^\ast})&= & \, \, 
\widetilde{D}({\bm e_j^\ast})D({{\bm e}_i}) \ .
\eeqa
The double commutant property implies  Morita equivalence as we shall show  in the 
section \ref{er}.

Alternatively the quantum tori relations $\mathbb T_\kappa$ and 
$(\mathbb T_\kappa)^\prime$ can be written as two commuting copies of  the $W_{\infty}$  algebra (see  \cite{DV93}):
%\marginpar{reference \cite{DV93} added}
%and $\tilde{ W}_\infty$, respectively
\beqa
\label{sine}
[D({\bm m}), D({\bm n})] &=& 
2i \sin \left(\pi \kappa (\bm m \times \bm n )\right)
 D({\bm m+n}) , \\[4pt]
[\widetilde{D}({\bm m^\ast}), \widetilde{D}({\bm n^\ast})]& =& 2i \sin \left(\frac{\pi}{\kappa} ({\bm m} \times {\bm n} )\right)
 \widetilde{D}({{\bm m}^\ast+{\bm n}^\ast}) .
\eeqa
We end up with a  splitting
$W_\infty \otimes \widetilde{ W}_\infty$ 
of the algebra of  ``zero modes'' of the Hamiltonian $H^\prime$ 
with a periodic potential into two commuting algebras
depending on the Bravais lattice of periods and its dual, respectively.

A word of caution: the algebras $W_\infty $ and $\widetilde{ W}_\infty$ both  commute with the algebra
of $a^{\pm}$ and the free Hamiltonian. Hence
they do not mix different Landau levels.
The operators of the algebra $\mathbb T_\kappa$ are exponentials of the ``zero mode'' operators
$b^{\pm}$; they create coherent states
 which are superposition of all states with different angular momenta
that live  in the lowest Landau level.
One can consider also the algebra of (momentum space) displacement operators
$
W_{k,\bar{k}}=\exp(\frac{k a^+ - \bar{k}a^-}{\sqrt{2}})
$
which will give rise to  another couple of $W_\infty$ algebras interpolating between different Landau levels.

\section{ Magnetic Translation Eigenfunctions }

Here we get % we are applying  the method of Choon-Lin Ho  in order to get 
the wavefunctions of Landau levels with quasiperiodic
conditions \cite{Ho}.
The  elements of the quantum tori $\mathbb T_\kappa$ and 
$(\mathbb T_\kappa)^\prime$ are expressed via the elementary
magnetic translations
% $D({\bm e_i})=D_i$ and   $\widetilde{D}({\bm e_i^\ast})=\widetilde{D}_i$
\beq
\label{Wil}
\ba{cccllc}
D({\bm m}) &=& q^{-\frac{1}{2}(m^1 m^2)} &
D({\bm e_1})^{m^1} D({\bm e_2})^{m^2} \qquad &,& {\bm m}= m^1 e_1 + m^2 e_2 ,  \\
\widetilde{D}({\bm n^\ast})&=& \tilde{q}^{-\frac{1}{2}(n_1 n_2)} &
 \widetilde{D}({\bm e_1^\ast})^{n_1}
 \widetilde{D}({\bm e_2^\ast})^{n_2}  \qquad & , & {\bm n}= n_1 e^{\ast 1} + n_2 e^{\ast 2} , 
\ea
\eeq
where the exponents  $m^i, n_i \in \mathbb Z$ are the winding numbers around the periods of the quantum  torus $\mathbb T_\kappa$
and its dual, respectively.
%\[
%D({\bm m})= q^{\frac{1}{2}(m^1 m^2)} 
%D({\bm e_1})^{m^1} D({\bm e_2})^{m^2} \quad \ ,
%\qquad 
%\widetilde{D}({\bm n^\ast})= \bar{q}^{\frac{1}{2}(n_1 n_2)} 
 %\widetilde{D}({\bm e_1^\ast})^{n_1}
 %\widetilde{D}({\bm e_2^\ast})^{n_2}  \ .
%\]
Every  element in $\mathbb T_\kappa$ commutes with every element
in $\mathbb T_{1/\kappa}$ since the magnetic translations
commute with the dual magnetic translations.

 When the flux density $\kappa$ is rational, meaning that some
rational number of quantized fluxes passes through a 
plaquette of the lattice $\Lambda$, the algebras
 $\mathbb T_\kappa$ and $\mathbb T_{1/\kappa}$ have a huge center,
and the algebras are  finite over their center.

When  the flux takes a rational value $\kappa = N/M$, the operators
$D({\bm{e_1}})$ and $D({\bm{e_2}})$
do not commute. However,
$D({\bm{e_1}})^M$ and $D({\bm{e_2}})^M$ do commute. Moreover
one has 
 $D(\bm{e_1})^M=\widetilde{D}({\bm{e_1^\ast}})^N$ and
 ${D}(\bm{e_2})^M=\widetilde{D}(\bm{e_2^\ast})^N$
hence we can choose a basis of  common eigenfunctions $\psi$
of (dual) magnetic translations 
\beq
\label{mn}
D(\bm{e_1})^M \psi = e^{i\alpha_1} \psi  
\left( = \psi \, \widetilde{D}(\bm{e_1}^\ast)^N  \right)
\quad , \quad D({\bm{e_2}})^M \psi = e^{i\alpha_2} \psi  \left(
=\psi \, \widetilde{D}(\bm{e_2}^\ast)^N \right)
\eeq
where the eigenvalues  $\alpha_i$
will be referred to as {\it vacuum angles}.
The common eigenfunctions $\psi$ transform as a left module of $\mathbb T_{\kappa}$ 
and  as  a right  module of 
${\mathbb T}_{\kappa^\prime}$.
The translations $D(\bm{e_1})^M$ and $D(\bm{e_2})^M$
as well as the operators $\widetilde{D}(\bm{e_2^\ast})^N $ 
and $\widetilde{D}(\bm{e_1^\ast})^N $ 
shift  on the bigger lattice 
\beq
\label{MN}
\Lambda_{MN}=\Z M e_1 \oplus \Z M e_2
= \Z N {e}_1^\ast \oplus \Z N {e}_2^\ast \ .
\eeq 
The lattice $\Lambda_{MN}$  generates the center of both the algebra $\mathbb T_\kappa$ 
and its dual $\mathbb T_{1/\kappa}$.
 The flux through an elementary plaquette of  $\Lambda_{MN}$
 is given by
\beq
\label{scales}
%L^2 Im \, \tau/(2\pi l^2_B)=
M^2 l^2 Im \, \tau/  (2 \pi l_B^2)=
M^2 \kappa= MN
. \eeq
With our assumption, $MN$ is  an integer
that equals the co-dimension of the center.

We will give explicitly
the eigenstates (\ref{mn}) of the (center of) the
magnetic translations and with their help construct the modules
of the quantum tori $\mathbb T_\kappa$ and $\mathbb T_{1/\kappa}$.
As a side corollary we get
\begin{prop} 
\label{dege}Let the parameter $\kappa$ be a rational number $\kappa = \frac{N}{M}$ with $M$ and $N$  coprimes, $gcd(M,N)=1$.
Then the degree of degeneracy of each Landau level is $MN$. 
\end{prop}

\noindent {\bf  Remark.}
The degree of degeneracy $MN$ is equal to the dimension of
the translation algebra
  $\mathbb T_\kappa \times \mathbb T_{1/\kappa }$, quotiented by its center generated in the lattice $\Lambda_{MN}$. The algebra $\mathbb T_\kappa \times \mathbb T_{1/\kappa }$ (considered
	as a discrete subgroup of all translations)
	is finite over its center
and its finite part is isomorphic to	
\[
\Lambda_{MN}^\ast/\Lambda_{MN}\cong \mathbb \Z_N \times \mathbb \Z_M \ .
\]

%In complex coordinates the

Magnetic  translation  of a wavefunction $\psi(z, \bar{z})= \langle z, \bar{z}| \psi \rangle $ Eq. (\ref{Du}), on the Bravais lattice for the complex coordinates $z=(x+iy)/l_B$ ({\it i.e.,} for the standard
complex structure)
looks like 
\beq
\label{mtr}
\ba{lcl}
D(\bm{e}_1) \psi (z, \bar{z})&=& e^\frac{ (z - \bar{z} )\l0}{{4  }} \, \psi (z - \l0, \bar{z}-\l0) \ ,
\\
D(\bm{e}_2) \psi (z, \bar{z}) &=& e^\frac{(\bar{\tau}z - \tau \bar{z})\l0 }{{4 }}\psi (z - \tau \l0 , \bar{z}-  \bar{\tau} \l0 ) \ ,
\ea
\eeq
where we use the dimensionless spacing $ \l0=l/l_B$ .

%\beq
%\label{mtr}
%\ba{lcl}
%D(\bm{e}_1) \psi (z, \bar{z})&=& e^\frac{l (z - \bar{z} )}{{4 \, Im \tau}} \, \psi (z - l, \bar{z}-l) \ ,
%\\
%D(\bm{e}_2) \psi (z, \bar{z}) &=& e^\frac{l (\bar{\tau}z - \tau \bar{z} )}{{4 \, Im \tau}}\psi (z - l \tau , \bar{z}- l \bar{\tau} ) \ .
%\ea
%\eeq

The magnetic translations $D(\bm{m})$ in the bigger lattice $\bm{m}\in \Z M\bm{e}_1+\Z M\bm{e}_2$
 has the eigenfunctions $\psi$
according to Eqs (\ref{mtr}) %leads to identities
\beqa
e^{-M (z-\bar{z})\l0/{4  }}\psi(z+M \l0,\bar{z}+M\l0)&=&
e^{i\alpha_1}\psi(z,\bar{z}) \\ 
e^{- M (\bar{\tau}z - \tau \bar{z} )\l0/{4 }}
\psi(z+\tau M \l0,\bar{z}+ \bar{\tau}M\l0)&=&
e^{i\alpha_2}\psi(z,\bar{z}) \ .
\eeqa
%\beqa
%e^{-M l_1(z-\bar{z})/{4 \, Im \, \tau }}\psi(z+M l_1,\bar{z}+Ml_1)&=&
%e^{i\alpha_1}\psi(z,\bar{z}) \\ 
%e^{- M l_2(\bar{\tau}z - \tau \bar{z} )/{4 \, Im \, \tau }}
%\psi(z+\tau M l_2,\bar{z}+ \bar{\tau}Ml_2)&=&
%e^{i\alpha_2}\psi(z,\bar{z}) \ .
%\eeqa

We now introduce a new variable $\z=z / Ml_0$.
%and new parameter $\mathcal{T}=\tau l_2/l_1$.
The wavefunction renamed after rescaling
  $\Psi(\z)=\psi(z / M\l0)$ 
is quasi-periodic with  conditions 
on periods $1$ and $\tau$ % $\mathcal{T}=\tau l_2/l_1$
\beqa
\Psi(\z+1,\bar{\z}+1)&=&
e^{i\alpha_1+ \pi MN(\z-\bar{\z})/{2 \, Im \, \tau}} 
\Psi(\z,\bar{\z})   \\
\Psi(\z+\tau,\bar{\z}+\bar{\tau})&=&
e^{i\alpha_2 +\pi MN(\bar{\Tau}\z-\Tau \bar{\z})/{2 \, Im \, \tau}}\Psi(\z,\bar{\z})
\eeqa
where  we have used\footnote{We could have used the complex structure $z=(x+ \tau y)/l_B$ 
to get the same result 
%Then the factor $Im \, \tau$ is present from the beginning in
through magnetic translations, Eq.(\ref{trance}).
% one has the factor $Im \, \tau$ from the start
 However  we have $M^2 l^2_0= 2\pi MN$
due to $%Im_{\J}\, \bm{e}_2= 
Im_{\J}\, \tau =1$  with respect the complex structure $\J$, i.e., the Bravais plaquette is the image of a square  in $(x,y)$-plane.} Eq. (\ref{scales}) so that  $M^2 l^2_0 Im \, \tau =M^2 2\pi \kappa= 2 \pi MN$.

The function $\Psi$ with the latter boundary conditions is determined by  a holomorphic function $F(\z)$  times
an appropriate factor $\eta(\z,\bar{\z})$ (such that we retrive
a periodic condition of $F(\z + 1) = F(\z)$ ) , 
\[
\psi(\z,\bar{\z})= \eta(\z,\bar{\z}) F(\z)
\qquad 
\]
In fact, the scale factor $\eta(\z,\bar{\z})$ which does the job reads %with the use of (\ref{scales}) takes the form 
\[
\eta(\z,\bar{\z})=
%\exp\left(\frac{M^2 l_1^2\z(\z-\bar{\z})}{4 Im \tau}
%+ i \alpha_1 \z\right) =
\exp\left(\frac{\pi MN \z(\z-\bar{\z})}{2 Im \, \tau}
+
i \alpha_1 \z\right) \ .
\]
The  boundary conditions on
the holomorphic part $F(\z)$ are periodic for the shift $\z \rightarrow \z + 1 $ % by construction $F(\z + 1) = F(\z)$
and quasiperiodic for $\z \rightarrow \z + \tau$ with period $ \tau  $:
\[
F(\z+1,\tau) = F(\z,\tau) \ ,
\qquad \qquad
F(\z+ \tau,\tau)= e^{-i \pi MN \tau} e^{-i 2 \pi MN (
\z +\gamma)} F(\z,\tau) 
\ .
\]
for $\gamma = \frac{\tau \alpha_1 -\alpha_2}{2 \pi M N}$
depending on the vacuum angles $\alpha_1$ and $\alpha_2$.

The quasi-periodic boundary conditions for   $F(z)$ are  to be compared  to the functional equations of the theta function $\vartheta(z, \tau)$ which is a quasiperiodic
function (holomorphic section) on the torus $\C/(\Z\oplus \tau\Z)$.

%These conditions are to be compared with the boundary conditions of
\begin{deff}
Theta function $\vartheta^K_{r}(z, \tau)$ of level $K\in\Z$
with modular parameter $\tau$ reads
\[
\vartheta^K_{r}(z, \tau)= \sum_{n\in \Z}
 e^{i\pi \tau K\left(n+\frac{r}{K}\right)^2}
e^{i 2 \pi K z \left(n+\frac{r}{K}\right)  } \qquad r \in \Z/K\Z:=\mathbb Z_K \ .
\]
%The index $r \in \mathbb Z_K$ 
\end{deff}
We compare the boundary conditions of our holomorphic  function $ F(\z, \tau)$
with the one of the theta function
 $\vartheta^K_{r}(z, \tau)$ of level $K=MN$ 
\beq
\label{quasip}
\vartheta^K_r(z+1,\tau) = \vartheta^K_r(z,\tau) \ ,
\qquad \qquad
\vartheta^K_r(z+ \tau,\tau)= e^{-i \pi K \tau} e^{-i 2 \pi K 
z } \vartheta^K_r(z,\tau) %\qquad \tau:=il_2/l_1 
\ .
\eeq
Hence we can identify\footnote{The solution has an additional freedom of multiplication by a function  $g(\tau)$ which is not function of $z$, $\partial g/\partial \z=0$, that is, $g(\tau) F(\z)$ will again satisfy the same functional equations as $F(\z)$} the function $F(\z)$
with the theta function $\vartheta^{MN}_r(\z+\gamma,\tau)$ of level $MN$. 
To summarize we have just proven  the following

\begin{prop} For the rational magnetic flux $\kappa=N/M$, 
the wavefunctions in the lowest Landau level 
 have  a finite basis of dimension $MN$. 
It is spanned by the  eigenfunctions of the magnetic translations
\[
\langle z,\bar{z}| j k\rangle= %\psi_{jk}(z,\bar{z})=
\Psi_{jk}(\z,\bar{\z}) , 
%\quad j=1, \ldots, M \quad k=1, \ldots, N
\qquad \z = \frac{z l_B}{Ml}
\] which are    coherent ground states.
%generating the lattice $\Lambda_{MN}$ (\ref{MN}). 
These $MN$ eigenfunctions are
%indexed by  $j=1, \ldots, M$ and $k=1, \ldots, N$,
indexed by
the  factor-lattice $\Lambda_{MN}^\ast/\Lambda_{MN}:
%\cong \mathbb \Z_N \times \mathbb \Z_M
$ 
\[
r_{jk}= jN + k M%\frac{jM+kN}{MN}  
%\frac{j}{M} +\frac{k}{N}
 \in  \Lambda_{MN}^\ast/\Lambda_{MN}\cong \mathbb \Z_M \times \mathbb \Z_N  %\qquad j=1, \ldots, M 
\ .
\]
 Given the vacuum angles $\alpha_1$ and $\alpha_2$,
the eigenfunctions of the magnetic translations (\ref{mn})
are amenable to $\vartheta$-functions of level $MN$
\beq
\label{psjk}
\Psi_{jk}(\z,\bar{\z}):=
\exp\left(\pi MN \frac{\z(\z-\bar{\z})}{2\, Im\, \tau}
+
i \alpha_1 \z\right)\vartheta^{MN}_{r_{jk}}\left({\z+\gamma}, \tau \right) 
\eeq
where  $\tau$ is the modular parameter of the compactified space
%configuration space torus
$T_\tau$ %$\mathcal{T} :=\tau l_2/l_1$
and the shift angle is 
$\gamma=\frac{ \tau \alpha_1- \alpha_2}{2 \pi MN}$.
\end{prop}

%{\bf Remark.} Note, that the the wavefunction $\Psi_{jk}(\z,\bar{\z})$ can be multiplied by an arbitrary function $\xi(\Tau)$.

Similarly at the $n$-th Landau level,
the wavefunctions % of energy $\hbar \omega(n+\demi)$
are obtained by the action of the raising operator $a^+$:
\[
\Psi^{(n)}_{jk}= a^{+ n}\Psi_{jk} \ .
%\bar{z}^n \langle z,\bar{z}| j k\rangle \qquad
%\Psi^{(0)} = \Psi
\]
Thus the degeneracy on any Landau level  is $MN$
which proves  Proposition \ref{dege}.

%\marginpar{about  Manin's quantum theta }
The states in any Landau level   are expressed in terms of  classical theta-functions.
The  operators behind this construction are the magnetic translation operators. The state-operator correspondence
has been used by  Manin in order to introduce the
 quantum theta-functions \cite{Manin},  these are operator-valued functions acting on the states-wavefunctions  
(spanning a module of the non-commutative torus).\footnote{ For nice and friendly for physicist introduction to
Manin's  quantum theta-functions based on $kq$-representation  see \cite{C-YK}}

\section{Modular Invariance and Coherent States}

In the Landau problem on toroidal geometry or rather
in the problem of Bloch electrons in a periodic potential
we encounter two different types of tori:  the  configuration space
torus $T_\tau$ is  a one-dimensional complex curve  whereas the  
magnetic translation operators close a non-commutative algebra
which is the quantum torus $\mathbb T_\kappa$  ( Eq. (\ref{qtor}) ).

The magnetic translation operators (\ref{Wil}) with loops  wrapped around the periods of $T_\tau$
 are the generators of $\mathbb T_\kappa$. These are the topologically nontrivial global gauge transformations.
The magnetic flux through a minimal loop, {\it i.e.,} through the lattice facette is  $\kappa$, depending
on the ratio $l/l_B$ and $Im \,\tau$, Eq. (\ref{kap}).
The  parameter $\tau$ of %the  configuration space 
$T_\tau$ encodes the ratio of the periods of the crystal lattice (and thus  the complex structure on $T_\tau$ which is scale invariant).

%The  path-integral formalism requires a functional
%on the Hilbert space.
%\[
%Z= \sum_{states} q^n \dim
%\]

The norm of a state  in a Landau level
is a complex function of the modular parameter $\Tau$. 
For the states in the lowest Landau level, Eq.(\ref{psjk}),  the norm reads
\beq
\label{psipsi}
\langle\Psi_{jk}|\Psi_{jk}\rangle(\Tau,\bar{\Tau})=
\int  \,   \Omega_{\Tau} \,
e^{-\frac{\pi MN }{2\, Im\, \Tau} |\z-\bar{\z}|^2}
e^{i \alpha_1 (\z-\bar{\z})}
\vartheta^{MN}_{r_{jk}}\left({\z+\gamma}, \Tau \right) 
\vartheta^{MN}_{r_{jk}}\left({\bar{\z}+ \bar{\gamma}}, 
\bar{\Tau} \right) 
\eeq
where the volume form is
 $\Omega_{\Tau}
= \pi MN {d\z \wedge d\bar{\z}}/{(\Tau -\bar{\Tau})}$.

We introduce a new shifted complex coordinate on the torus\footnote{
By some abuse we denote it again by $z$.
The old $z=(x+\tau y)/l_B$ and the new $z$
are related by a shift and scaling $\frac{z}{Ml}+\gamma \rightarrow \z+\gamma \rightarrow z$ thus the complex structure is not altered.
}, $z=\z+\gamma$ and 
obtain after completing the square\footnote{Since the vacuum angle $\alpha_1= i  \pi M N (\gamma-\bar{\gamma})/Im \, \Tau$ we can perform a global gauge transformation of the wavefunction
 $\Psi_{jk}(\z,\bar{\z})\rightarrow
\Psi_{jk}(\z,\bar{\z})
e^{i\pi MN \frac{\gamma(\gamma-\bar{\gamma})}{ ( \Tau-\bar{\Tau})}}
$.
}
a Gaussian
measure on the space of holomorphic and anti-holomorphic functions:
\beq
\label{ps2}
\langle\Psi_{jk}|\Psi_{jk}\rangle(\Tau,\bar{\Tau})=
\int   \frac{dz \wedge d\bar{z}}{ \Tau- \bar{\Tau}}% \,   \Omega_{\Tau} \,
\exp\left({-\frac{\pi MN }{2\, Im\, \Tau} |z-\bar{z}|^2}\right)
\vartheta^{MN}_{r_{jk}}\left({z}, \Tau \right) 
\vartheta^{MN}_{r_{jk}}\left({\bar{z}}, 
\bar{\Tau} \right)  .
\eeq
{\bf $z$-invariance.}
The above integrand 
is invariant under a  translation of the $z$-variable with the 
periods of the torus.
Indeed the translation
invariance 
is obvious for the period  $z \rightarrow z +1$.
To show the invariance 
for the other period $z \rightarrow z +\Tau$, we note 
\beqa
&&
\vartheta^{MN}_{r_{jk}}\left({z} +\tau, \Tau \right) 
\vartheta^{MN}_{r_{jk}}\left({\bar{z}}+\bar{\tau}, 
\bar{\Tau} \right) \exp\left({\frac{i\pi MN }{ \tau-\bar{\tau}} (z-\bar{z}+\tau-\bar{\tau})^2}\right)\\
&=&
\vartheta^{MN}_{r_{jk}}\left({z} , \Tau \right) 
\vartheta^{MN}_{r_{jk}}\left({\bar{z}}, 
\bar{\Tau} \right) \exp\left({\frac{i\pi MN }{ \Tau- \bar{\Tau}} (z-\bar{z})^2}\right),
\eeqa
where one has to make use of the quasiperiodicity  of
the $\vartheta^K(z,\Tau)$, Eq.(\ref{quasip}).
The choice of the non-holomorphic prefactor is crucial for the periodicity. This is a translationally invariant gauge.

Hence the states are indeed well defined on the factor space
of the complex plane
\[
T_\tau= \C/ (\Z +\tau \Z) \ .
\]

\noindent
{\bf  Modular transformations.}
There are different values of $\Tau$ for which the tori $T_\Tau$ are isomorphic as complex manifolds. These are precisely the tori whose  parameters are  related by the modular transformation
\[
 T_\Tau \cong T_{\Tau^\prime} 
\qquad  \Leftrightarrow \qquad 
\Tau^\prime= \frac{a \Tau + b}{c \Tau + d}  \quad \mbox{with}\quad\left(\ba{cc}a & b\\ c & d \ea \right)  \in SL(2,\Z) \ .
\]
A modular transformation changes the basis of the lattice
$\Lambda=\Z \oplus \Tau \Z$,  thus keeping the quotient invariant,
$T_\tau=\C/\Lambda$.
However, a choice of the basis  of $\Lambda$  fixes
the periods and hence the holonomy loops for the 
magnetic translation operators providing the basis of
the non-commutative torus $\mathbb T_\kappa$.
Therefore,
the moduli space of the tori is the space of parameters $\Tau$ modulo the action of the modular group $SL(2,\Z)$.
The transformations $T: \Tau \rightarrow \Tau^\prime=\Tau +1$ and 
$S: \Tau \rightarrow \Tau^\prime=-1/\Tau $ are in $SL(2,\Z)$,
and correspond to matrices
\[
T= \left(\ba{cc} 1 & 1\\ 0& 1 \ea \right) \quad , \quad
S= \left(\ba{cc} 0 & 1\\ -1& 0 \ea \right)  . 
\]
It is not difficult to see that $S$ and $T$ generate the modular group $SL(2,\Z)$.
The fundamental domain  for the $SL(2,\Z)$-orbits
% is parametrized by the non-equivalent
%complex structures on torus. 
 can be chosen to be the domain $\tau \in \C $
\[
\left\{  Im \, \Tau>0 \right\} \qquad \mbox{and}\qquad  \left\{ \ba{ccrcl}
|\Tau| >1  && - 1/2 <  &Re \, \Tau& < 0  \\
|\Tau| \geq 1  && 0 \leq & Re \, \Tau& \leq 1/2
\ea \right.
 \ .
\]
%\[
%\left\{ Im \, \Tau>0 \right\} \cap (\left\{
%|\Tau| >1  \quad - 1/2 < Re \, \Tau < 0 \right\} \cup
%\left\{  
%|\Tau| \geq 1  \quad 0 \leq Re \, \Tau \leq 1/2 \right\}) 
    %\ .
%\]
Every $SL(2,\Z)$-orbit has one point in  the fundamental domain.
This is the moduli space of the complex structures on
the  torus. More on the modular invariance in the context of
conformal field theory can be found in the lectures 
\cite{NT}.

%the interiour of fundamental domain (or possible two points if it is on the boundary).

\noindent
{\bf Partition function.}
In conformal field theory %the theory on the torus
we have path-integral formalism for any Riemann surface.
In particular for the Riemann surface of genus 1 and modulus $\tau$, that is, 
the torus $T_\tau$,  the {\it partition function} $Z(\tau)$  is the vacuum functional
\[
Z(\tau) = Tr (q^{L_0-\frac{c}{24}}\bar{q}^{\bar{L}_0-\frac{c}{24}})
\qquad q=e^{i 2\pi \tau}
\]
where $L_0$ and $(\bar{L}_0)$ are the holomorphic and antiholomorphic
conformal energy operators. The constant $c$ is the central charge
of the conformal theory. The characters of the
chiral affine algebra are
\[
\chi_\mu(\tau) = Tr (q^{L_0-\frac{c}{24}}) \qquad q=e^{i 2\pi \tau}
\]
and they carry representation of the modular group $SL(2,\Z)$.
%The partition function $Z(\tau)$ is expressible as bilinear combination of character of affine algebras.
The  partition function $Z(\tau)$ is a sesquilinear pairing of characters with non-negative integer coefficients $Z_{\mu \nu}$:
\[
Z(\tau)=\sum_{\mu}
\chi^\ast_\mu Z_{\mu \nu} %(z,\tau) 
\chi_\nu \quad ,  %(z, \tau) 
\quad Z_{\mu \nu} \in \Z_{\geq 0}  \ .
\]
The integer matrices $Z_{\mu \nu}$ satisfying the 
$SL(2,\Z)$-invariance conditions 
\[
SZ=ZS \qquad TZ=ZT
\]
give rise to  modular invariant partition functions $Z(\tau)$.

A free boson on the torus has partition function
\[
Z_{boson}(\tau) = \frac{1}{\sqrt{Im \, \tau}} \frac{1}{|\eta(\tau)|^2}
\]
where the Dedekind  function $\eta(\Tau)$ is given as the infinite product
\[
\eta(\Tau)= q^{1/24} \prod_{n=1}^\infty (1 - q^n) \qquad q=e^{2 \pi \Tau} \ .
\]
One can check its modular  properties under $T$ and $S$ transformations:
\[
\eta(\Tau+1) = e^{i \pi /12} \eta(\Tau) \qquad ,
\qquad \eta(-{1}/{\Tau}) = \sqrt{-i \Tau} \eta (\Tau) \ .
\]
The latter transformations exactly compensate the transformations of $Im \, \tau$ and thus
 guarantee that the partition
function of the free boson  $Z_{boson}(\tau)$ is modular invariant.

The  affine  lattice characters are written as the ratio of  theta functions
of level $K$  and the Dedekind function
\[
\chi_\mu (z,\tau) = \vartheta_\mu^K(z,\tau)/ \eta(\tau)
\quad, \quad  \mu \in \Lambda^\ast_K / \Lambda_K \ .
\] 
The role of $\eta(\tau)$ is again to compensate for the nontrivial
modular transformations of the theta functions, which are
modular half-forms, just as the Dedekind function.

The modular transformations of the affine characters are 
remarkably beautiful \cite{Kac}: 
\beqa
\label{t1}
\chi_\mu (z,\tau +1) &=& e^{2 \pi i \left(\frac{|\mu|}{2}^2- \frac{1}{24} \right)} 
\chi_{\mu} (z, \tau) \ ,\\
\label{s1}
e^{-i \pi  K z^2/{\tau}}\chi_\mu(\frac{z}{\tau},-\frac{1}{\tau})
&=& |\Lambda_{K}^\ast/\Lambda_{K}|^{-\demi} 
\sum_{\mu^\prime \in\Lambda^\ast_K / \Lambda_K} e^{- i 2\pi (\mu|\mu^\prime)} \chi_{\mu^\prime}(z,\tau) \ .
\eeqa
Thus the partition function is a sum over the Hilbert space of states
with weights depending on the energy.  We sum first on states of a given Landau level and then sum with weights over all levels.
As the degeneracy of each Landau level is the same, we expect to
have factorization of the partition function.
All the physically meaningful quantities in a theory defined on a torus
should be  modular invariant.
\begin{lemma}
The sum of the norms of all  states
%$\xi_{jk}(\z,\bar{\z},\Tau)$  
in the lowest Landau level normalized by  the Dedekind function $\eta(\Tau)$
is a modular invariant
\[
 \tilde{Z} (\Tau)
%=
%\sum_{{(j,k)}\in \Z_M\times \Z_N} |\xi_{jk}(\Tau)|^2
= \sum_{r_{jk}\in \Z_M\times \Z_N} \frac{\langle\Psi_{jk}^{}|\Psi_{jk}^{}\rangle}{|\eta(\Tau)|^2} \ .
\]
\end{lemma}
{\bf Proof.}
The summation on the weights $\mu={r_{jk}}
\in \Lambda_{MN}^\ast/\Lambda_{MN}\cong \mathbb \Z_M \times \mathbb \Z_N$ yields
\beq
\label{ps3}
\tilde{Z} (\Tau)=
\int   \frac{dz \wedge d\bar{z}}{ \Tau- \bar{\Tau}}% \,   \Omega_{\Tau} \,
\exp\left({-\frac{\pi MN }{2\, Im\, \Tau} |z-\bar{z}|^2}\right)
\sum_{\mu \in \Lambda_{MN}^\ast/\Lambda_{MN} }
\chi_\mu^\ast\left({{z}}, 
{\Tau} \right) \chi_\mu\left({z}, \Tau \right)  \ .
\eeq
The modular invariance $\tilde{Z}(\tau+1)=\tilde{Z}(\tau)$ 
%$T: \tau \rightarrow \tau +1$ 
is straightforward due to the compensation of the phase factor in Eq.(\ref{t1}). The invariance $\tilde{Z}(- 1/ \tau)= \tilde{Z}(\tau)$
follows from Eq.(\ref{s1}) and the orthogonality relations
\[
\delta_{\mu^\prime \mu^{\prime\prime}}
= (MN)^{-1} \sum_{\mu \in \Lambda_{MN}^\ast/\Lambda_{MN}} e^{- 2 \pi i (\mu^\prime|\mu)} 
e^{2 \pi i (\mu|\mu^{\prime \prime})} \ .
\]
We conclude that $Z(\tau)$ is a good candidate for the
partition function on the torus.

\section{Metaplectic Representations and Squeezed States}
 
The Heisenberg algebra $\heis_3$ is realized as 
the algebra of  creation and annihilation operators
$b^\pm$ with the oscillator relation $[b^-,b^+]=1$.
The three quadratic symmetric  polynomials in  $b^+$ and $b^-$
\beq
\label{pb1}
J_\pm = \frac{1}{4} \{b^\pm, b^\pm \} \ ,\qquad
J_0= \frac{1}{4}\{b^+, b^- \}
\eeq
close an algebra $\mathfrak{sp}(2,\R)\cong \mathfrak{sl}(2,\R)
\cong \mathfrak{su}(1,1)$:
\beq
\label{pb2}
[J_0, J_{\pm}]=\pm J_{\pm} \ ,\qquad\qquad[J_+,J_-]=-2J_0
\eeq
which is known as the metaplectic representation of 
$\mathfrak{sl}(2,\R)$.
The action of quadratic generators %operators 
$J_\pm$ and $J_0$ %algebra $\mathfrak{sl}(2,\R)$ 
%acts
 on the generators $b^\pm$
\beq
\label{mp1}
[J_0, b^\pm] = \pm \demi b^\pm
\qquad
[J_\pm, b^\mp]=\mp b^\pm \qquad
[J_\pm, b^\pm]=0
\eeq
gives the algebra of automorphisms of $\heis_3$.
Hence the quadratic and the linear polynomials in $b^+$, $b^-$ together with
the identity operator $1$ close an algebra which is a semidirect product $\heis_3 \rtimes
\mathfrak{sl}(2,\R) $.

When exponentiated the metaplectic representation (\ref{mp1}) 
is double valued, it gives rise to a double covering
of the group $SL(2, \R)$ which is referred to as the {\it metaplectic representation} $Mp(2,\R)$.
The automorphisms of the Heisenberg group  $H_3$ are
given by the metaplectic group $Mp(2,\R)$. The
quantum torus or equivalenly the algebra of the displacement operators $D(\bm{u})$ provides a representation of the Heisenberg group $H_3$ which also carries an action of $Mp(2,\R)$.

We now describe the semi-direct group product $H_3\rtimes Mp(2,\R)$.
We can either work with the double cover of $SL(2,\R)$  
or with $SU(1,1)$.
The Cartan operator $2J_0=\frac{1}{2} \{b^-, b^+\}
\in \mathfrak{su}(1,1)$. The operator $2J_0=(N+\demi)$ with
integer spectrum
is exponentiated to the unitary operator $U_\phi = \exp(i\phi 2J_0)$ so that
 \[
U_\phi \left(\ba{c}
b^+\\
b^-
\ea\right) U_\phi^{-1} = e^{\pm i{\phi}}\left(\ba{c}
b^+\\
b^-
\ea\right) . 
\]
It might come as a surprise that the {\it squeezing operators} which are massively
used in quantum optics
\beq
\label{squeeze}
S(w) = \exp \demi\left( w (b^{+})^2 - \bar{w}(b^-)^2 \right) \quad ,
\quad w=re^{i\varphi}
\eeq
 belong to the metaplectic representation of 
$SU(1,1)$.

The operator $S(w)$ acting 
on the creation and annihilation operators $b^\pm$
 maps them to another pair $b^\pm_\tau$ given by
\beq
\left(\ba{c}
b^+_ \tau\\
b^-_\tau
\ea\right) = S(w) \left(\ba{c}
b^+\\
b^-
\ea\right) S^{-1}(w) =\left(\ba{rr} \cosh r &e^{-i\varphi} \sinh r\\ 
e^{i\varphi} \sinh r & \cosh r \ea\right) 
\left(\ba{c}
b^+\\
b^-
\ea\right)  . 
\eeq
Thus the squeezing operator $S(w)$ is intertwining different representations of $\mathfrak{h}_3$. %the oscillator representation. 
Such intertwining of oscillator representations
is also known as
 Bogoliubov transformations in e.g., superfluidity theory.
The $SU(1,1)$-symmetry applies to the geometry of the Laughlin Fractional Quantum Hall states in the framework of  Chern-Simons matrix model  see \cite{Lapa}.
%\marginpar{new very interesting reference, see eq 2.22 !!! for the approach of Haldane to LaughlinFQHE, it might be that changing the complex structure  locally is exactly what is needed }
%In order to get the dependence between the squeezing parameter $w$
%and $\tau$ we are going to 
In other words the squeezing operator $S(w)$ interpolates between different  complex structures, thus
perturbing the standard complex structure
 $\J_0$  to a new one $\J(\tau)$
\[
S(w)\J_0 S^{-1}(w)=
i \left(\ba{rr} \cosh 2r &- e^{i\varphi} \sinh 2r\\ 
e^{-i\varphi} \sinh 2r & -\cosh 2r \ea\right) =\J(\tau)
\]
\[ = 
\left(\ba{rr} \cosh r &e^{-i\varphi} \sinh r\\ 
e^{i\varphi} \sinh r & \cosh r \ea\right)
\left(
\ba{cc} i&0\\ 0& -i
\ea\right)
\left(\ba{rr} \cosh r &-e^{-i\varphi} \sinh r\\ 
-e^{i\varphi} \sinh r & \cosh r \ea\right) . 
\]
The ``perturbed" complex structure is to be compared to 
the general complex structure $\J$, (given by Eq. (\ref{jj})
but in the $(x,y)$-basis) dependent on 
the modulus $\tau$:
\beq
\J(\tau)=
 \frac{i}{2 \, Im \, \tau} \left(\ba{cc} 1 + |\tau|^2 &
- 2 i Re + (1-|\tau|^2) \\ 
- 2 i Re - (1-|\tau|^2) & -1 -| \tau|^2 \ea\right)
\eeq
Comparing the matrices above  one determines the (``squeezing'') parameters $w=w(\tau)=r(\tau)e^{i\phi(\tau)}$ as functions of the complex structure modulus $\tau$
\[
\cosh 2r = \frac{1+|\tau|^2}{2 \, Im \tau} \ ,
\qquad \sinh 2r \cos \varphi = \frac{1-|\tau|^2}{2 \, Im \tau}
\ ,
\qquad
\sinh 2r \sin \varphi = - \frac{Re \, \tau}{Im \, \tau} \ .
\]
Thus we get the dependence
\[
\tan \varphi = \frac{Re \tau }{|\tau|^2-1} \ ,
\qquad \sinh 2r = \frac{1-|\tau|^2}{2\, Im \, \tau}\sqrt{1+\tan^2 \varphi} \ .
\]
In particular, when $w$ is real we have $Re \, \tau =0$
and  ``squeezing'' of the $Im\, \tau$
\[ \varphi =0 \ ,  \qquad Re \, \tau =0 \ , \qquad
e^{-2r}= Im \, \tau \ .\]
Roughly speaking, in the plane geometry the operator $S(w)$ acting on the complex structure $\J_0$ is relating  it to any other $\J$.
For the compactified theory on the torus different  complex structures are interpolated by
a metaplectic representation of the modular group $SL(2,\Z)$,
the group of invariance of the Bravais lattice $\Lambda$.
 
The argument $z=x +\tau y$ of the theta function carries
a representation the Heisenberg group $H_3$ of the magnetic translations, while the argument $\tau$ is transforming in the metaplectic group $Mp(2,\R)$.
It was Andr\'e Weil who revealed that
the theta function $\vartheta(z,\tau)$ is a matrix representation 
of the cross-product $H_3 \ltimes Mp(2,\Z)$,  \cite{Tata}.

\section{ Matrix algebras and Morita equivalence}
\label{er}
Magnetic translations commute with the Hamiltonian of the  Landau problem 
hence do not alter the energy.
The lowest Landau level consisting of  the ground states in Hilbert space  is a  representations
of both the magnetic algebra $\mathbb T_\kappa$ and the dual magnetic algebra
$\mathbb T_{\kappa^\prime}$. In fact any Landau level 
carries one and the same  structure of a module of
two non-commutative tori whose magnetic flux parameters are related by 
${\kappa}^\prime = 1/\kappa$.

\begin{lemma} Let us consider a rational magnetic flux $\kappa=N/M$.
 The wave functions  $\Psi_{jk}(z,\bar{z})= \langle z,\bar{z}| j k\rangle$, $j\in \mathbb Z_N$ and $k\in \mathbb Z_M$ span a left
module of  the quantum torus $\mathbb T_\kappa$ and a right
module of the  dual quantum torus
$\mathbb T_{\kappa^\prime}$
\[
\ba{lcl}
 D({\bm{e_1}})  \Psi_{jk}%| j k\rangle 
= e^{i(\alpha_1 + 2 \pi j N)/M}\Psi_{jk} \ ,
 && \Psi_{jk} \widetilde{D}({\bm{e^\ast_1}}) % | j k\rangle 
= e^{i\alpha_2/N } \Psi_{j \, k-1} \ ,
\\
D({\bm{e_2}})\Psi_{jk}   
 =   e^{i\alpha_2/M }  \Psi_{j-1\,k}  \ ,
 &\qquad& 
 \Psi_{jk} \widetilde{D}({\bm{e_2^\ast}}) %| j k\rangle 
= 
e^{i(\alpha_1 + 2 \pi k M)/N} \Psi_{jk} \ .
 \ea
\]
\end{lemma}

 The operator $D({\bm{e}_1})$ acts on the left module
by the $M\times M$-matrix
\[
D({\bm{e}_1}) = 
e^{i\alpha_1/M}\left(\ba{cccc}
1 & & &\\& e^{i 2 \pi N/M} & & \\
& & \ddots &\\
& & & e^{i 2 \pi (M-1) N/M}
\ea\right) 
\] where we have written only the nonzero entires.
Similarly $D({\bm{e}_2})$ acts by
\[
D({\bm{e}_2}) = 
e^{i\alpha_2/M}\left(\ba{cccc}
0 & &  &1\\1& 0&  &  \\
& \ddots& \ddots &\\
%&1 & 0& 
%\\
& & 1& 0
\ea\right)  \ .
\]
Note that both matrices are of trace zero (on the diagonal of the first matrix
we have sum of roots of unity, $\sum_{j=0}^{M-1} \omega^j=0$ when $\omega^M=1$).

\begin{deff} Let $E$ be a $(A,B)$-bimodule. The algebra $A$
is Morita equivalent to the algebra $B$ if there exists
a $(B,A)$-bimodule $F$ such that the following isomorphisms hold true:
\[
E\otimes_{B}F \cong A \qquad \mbox{as} \qquad A\mbox{-bimodule} \ ,
\]
\[
F\otimes_{A}E \cong B \qquad \mbox{as} \qquad B\mbox{-bimodule} \ .
\]
\end{deff}
In fact, 
here we encounter an example of $T$-duality
 which is the physical jargon for what the mathematicians  call Morita equivalence \cite{Schwarz98}.

\begin{prop} Let the flux $\kappa$ be a rational number $\kappa=N/M$. Then any $MN$-fold degenerate Landau level is  spanned by the  wavefunctions $\Psi_{jk}$ (\ref{psjk}).
The $M\times N$ matrix $$\Psi=\left\{\Psi_{jk}, j\in \mathbb Z_M, k\in \mathbb Z_N \right\}\qquad \mbox{is  a} \quad 
(\mathbb T_\kappa, \mathbb T_{\kappa^{\prime}})-\mbox{bimodule}$$
whereas 
the $N\times M$ matrix
$$\bar{\Psi}=\left\{\bar{\Psi}_{kj}, k\in \mathbb Z_N , j\in \mathbb Z_M \right\}\qquad \mbox{is  a} \quad (\mathbb T_{\kappa^\prime}, \mathbb T_{\kappa})-\mbox{bimodule} \ .$$
The bimodules $\Psi$ and $\bar{\Psi}$ provide the Morita equivalence
of the quantum torus $T_{\kappa}$ and its dual $\mathbb T_{\kappa^\prime}$ with $\kappa^\prime = 1/ \kappa$.
\end{prop}

To summarize, the theta functions exhibit a bimodule structure which
guarantees the $T$-duality of the algebra of magnetic translations
$\mathbb T_{\kappa}$ and its dual $\mathbb T_{\kappa^\prime}$.

Finally we consider he universal quantum enveloping algebra $U_q (sl_2)$   generated by
\beq
[J_3, J_\pm]=\pm J_\pm \ , \qquad [J_+,J_-]= \frac{q^{2J_3}-q^{-2J_3}}{q-q^{-1}} \ ,
\eeq
that can be
built from  the magnetic translation operators
\cite{Kogan, Ho} along the light-cone directions on the lattice.
Indeed it is straightforward to check
that the following combinations realize the algebra $U_q (sl_2)$:
\[
J_\pm:= \frac{D({\pm\bm{e}_1 \pm \bm{e}_2})-D({\mp\bm{e}_1 \pm \bm{e}_2})}{q-q^{-1}} \ ,
\qquad
q^{\pm J_3}:=D({\pm \bm{e}_1} ) \ , \quad 
\mbox{with} \quad  q=e^{i 2 \pi \kappa} .
\]
%with $q=e^{i 2 \pi \kappa}$. 
On the other hand another 
``dual'' copy of  $U_{\widetilde{q}} (sl_2)$ algebra for the dual parameter $\widetilde{q}=e^{i 2 \pi/ \kappa}$ is built
through the dual magnetic translations
\[
\widetilde{J}_\pm:= \frac{\widetilde{D}({\pm \bm e^\ast_1 \pm\bm e^\ast_2})
-\widetilde{D}({\mp \bm e^\ast_1
 \pm \bm e^\ast_2})}{\widetilde{q}-\widetilde{q}^{-1}} \ , 
\qquad
\widetilde{q}^{\pm J_3}:=\widetilde{D}({\pm \bm e^\ast_1}) \ , \quad 
\mbox{with} \quad  \widetilde{q}=e^{i 2 \pi/ \kappa} .
\]

The Bloch waves $\Psi^{(n)}_{jk}=a^{+n}\Psi_{jk}$ given by Eq. (\ref{psjk}), living at  the $n$-th Landau level, are entries of a    $M\times N$ matrix carrying 
a  left $M$-dimensional $U_q (sl_2)$-module and right 
$N$-dimensional $U_{\tilde{q}} (sl_2)$-module \cite{Ho}.

\section{Conclusion and Outlook}
From the geometric perspective,   theta-functions arise as sections of a bundle over the
torus of the compactified coordinate space in the presence of
a periodic potential. Applying the non-commutative geometry
paradigm we replace vector bundles with
modules of non-commutative algebras, hence 
the Bloch waves for electrons without spin are traded as modules of 
 Heisenberg-Weyl algebras of magnetic translations.
These modules provide a textbook example of Morita equivalence.

Next natural step is to consider Bloch waves for the electron
with spin.
The geometric point of view has the merit that  the spin interaction of the electron  
with the magnetic field is incorporated into  fermionic degrees of freedom living  on  spinorial bundles over the torus.  The super-symmetric counterparts of the  magnetic translations operators  will satisfy the super-sine algebra
\cite{DV92,DV93}, that is, the super-symmetric version of $W_\infty$
(\ref{sine}) which we can refer to as a non-commutative super-torus.
We expect that Bloch waves with spin can be expressed as  appropriate  supersymmetric theta-functions, 
%and these will play the role of  modules of the non-commutative super tori. 
the super-holomorphic sections  transforming
into a module of
pair of Morita equivalent non-commutative  super-tori.

The maximal symmetry  of the Landau problem is given by the symplectic morphisms $Sp(4,\R)$ of the phase space $(\R^4, \Omega)$. In view of the isomorphism $Sp(4,\R)\cong SO(2,3)$ it coincides with the $2+1$  Minkowski space conformal symmetry $SO(2,3)$ 
which in turn has a natural description in terms of the Jordan algebra of $2 \times 2$ symmetric real matrices. 
So in the present paper we have  prepared the ground for  treating the  symmetries of the Landau problem in the framework of Jordan algebras.

\section{Acknowledgement}

\noindent  One of us (TP)  is supported by TUBITAK through a BIDEP-2230 Visiting Scholar Grant and by
the Bulgarian National Science Fund Research Grant DN 18/3.
We also thank TUBA for partial financial support.
We are grateful to Ko\c{c} University for hospitality, especially the Office of VPRD (Vice President for Research and Development)
for support that came at  the right times.

\end{document}